\documentclass[twocolumn,showpacs,preprintnumbers,amssymb]{revtex4}

\usepackage{graphicx,epsf, epsfig, amssymb}% Include figure files
\usepackage{bm}
\usepackage{longtable}
\usepackage{color}

\def\be{\begin{equation}}
\def\ee{\end{equation}}
\def\beq{\begin{eqnarray}}
\def\eeq{\end{eqnarray}}

\begin{document}

\title{Comment on ``Kerr Black Holes as Particle Accelerators to Arbitrarily High Energy''}


\author{
Emanuele Berti$^{1,2}$, %\footnote{Electronic address: berti@phy.olemiss.edu}$,
Vitor Cardoso$^{2,3}$, %\footnote{Electronic address: vitor.cardoso@ist.utl.pt}$, 
Leonardo Gualtieri$^{4}$, %\footnote{Electronic address: Leonardo.Gualtieri@roma1.infn.it}
Frans Pretorius$^{5}$, %\footnote{Electronic address: fpretori@princeton.edu}
Ulrich Sperhake$^{1}$ %\footnote{Electronic address: sperhake@tapir.caltech.edu>}$, 
}

\affiliation{${^1}$ California Institute of Technology, Pasadena, CA 91109, USA}

\affiliation{${^2}$ Department of Physics and Astronomy, The University of
Mississippi, University, MS 38677, USA}

\affiliation{${^3}$ CENTRA,~Departamento.~de F\'{\i}sica,~Instituto.~Superior.~T\'ecnico, Av.~Rovisco Pais 1, 1049 Lisboa, Portugal}

\affiliation{${^4}$ Dipartimento di
  Fisica, Universit\`a di Roma ``Sapienza'' \& Sezione INFN Roma1, P.A. Moro 5, 00185, Roma, Italy}

\affiliation{${^5}$ Department of Physics, Princeton University, Princeton, NJ 08544, USA}

\begin{abstract}
It has been suggested that rotating black holes could serve as particle
colliders with arbitrarily high center-of-mass energy. Astrophysical
limitations on the maximal spin, back-reaction effects and sensitivity to
the initial conditions impose severe limits on the likelihood of such
collisions.
\end{abstract}

\maketitle
Ba\~nados, Silk and West (henceforth BSW) \cite{Banados:2009pr} recently
observed that collisions of point particles falling from rest into rotating
(Kerr) black holes (BHs) may have arbitrarily large center-of-mass (CM)
energies close to the event horizon if the BH is maximally spinning and one of
the particles has orbital angular momentum close to the value $L=L_{\rm
  scat}$ corresponding to marginally bound geodesics.

An important practical limitation on the achievable CM energies occurs
because, as pointed out by Thorne \cite{Thorne:1974ve}, the dimensionless spin
of astrophysical BHs should not exceed $a/M=0.998$. For $a/M=0.998$, Eq.~(14)
in BSW yields maximum CM energies of about $10\mu $ for particles of rest mass
$\mu$.
%well below the limits 
%achievable by particle accelerators. 
%{\bf For a 1TeV dark matter particle this yields 10 TeV, larger than what
%  current accelerators can produce, but still well below the limits achievable
%  in the collisions between cosmic rays and the Earth's atmosphere and crust.}

Even in the idealized scenario of {\it being given} an extremal BH,
back-reaction effects make high-energy scattering very unlikely.
%perhaps even less relevant than Hawking radiation.  
Neglecting gravitational radiation, upon absorption of a pair of colliding
particles of mass $\mu$ the dimensionless spin must be reduced by
$\epsilon\equiv 1-a/M\sim \mu/M$. After this first collision, Eq.~(14) in BSW predicts
that the new maximum allowed CM energy would be
%
%\be
$E_{\rm CM} \lesssim 10^{12}\left(\mu/1{\rm MeV}\right)^{3/4}\left(M/100{\rm
  M_{\odot}}\right)^{1/4}\,{\rm GeV}$,
%\nonumber\,,
%\ee
%
orders of magnitude below the Planck scale for typical values of the
parameters. The collision of (say) a {\em single}
electron pair would reduce the spin of a $100M_{\odot}$ BH enough to inhibit
any further Planck-scale collisions. A hypothetical dark matter particle
would need a mass $\mu\gtrsim 10^3$~TeV to allow for more than one 
Planck-scale event.

\begin{figure}[t]
\begin{center}
\epsfig{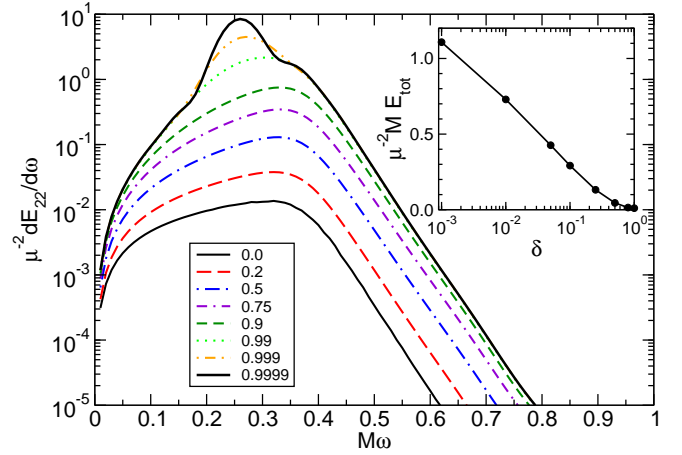}
\end{center}
\caption{\label{bump} Energy spectrum for $l=m=2$ and different values of
  $L/L_{\rm scat}$ (as indicated in the legend). Inset: for small $\delta$, a
  fit of the numerics yields $\mu^{-2}M\, E_{\rm tot}\sim
  -0.11-0.18\log\delta$.}
\end{figure}

The estimates in BSW neglect gravitational radiation, which will significantly
affect geodesics with $\delta=1-L/L_{\rm scat}\ll 1$ (notice that in the {\it
  critical} case $L=L_{\rm scat}$ it takes an infinite amount of proper time
for a particle to reach the horizon).  For small $\delta$, the
particle orbits around the marginally bound circular geodesic of a
Schwarzschild BH $N\simeq -(\sqrt{2}\pi)^{-1}\log\delta$ times.  For
near-extremal Kerr BHs we get $N\sim
-(2\pi\sqrt{2\epsilon})^{-1}\log(8\sqrt{\epsilon}\,\delta)$.
% $N\simeq
%\left(\Omega_{\rm mb}\dot{t}_{\rm mb}\,\log{k\delta}\right)/
%\sqrt{2\pi^2V''}$ times before plunging. Here $V$ is the effective radial
%potential, with $r^5V''=24M(L-aE)^2+6r\left(a^2(E^2-1)-L^2\right)+4Mr^2$,
%$\Omega\equiv d\phi/dt$ is the angular velocity, dots stand for derivatives
%with respect to proper time and $k=k(a)$ is a constant.  All quantities are
%evaluated at $r=r_{\rm mb}=2M\mp a+2M\sqrt{1\mp a/M}$ (for $E=1$). If $E=1$ and $a=0$, then $\Omega_{\rm
%  mb}=(8M)^{-1}$ and $N\sim (\sqrt{2}\pi)^{-1}\log\delta$.  For near-extremal
%Kerr BHs ($\epsilon=1-a/M\ll 1$) we get $N\sim
%(2\pi\sqrt{2\epsilon})^{-1}\log(8\sqrt{\epsilon}\,\delta)$.
This simple analysis suggests that for $\delta \ll 1$ the radiation should be
peaked at frequencies corresponding to marginally bound quasi-circular orbits
with orbital frequency $\Omega_{\rm mb}$ and that the total radiated energy
$E_{\rm tot}\sim N\sim -\log{\delta}$.

In Fig.~\ref{bump} we confirm these conclusions by a numerical calculation of
the energy radiated by particles of mass $\mu\ll M$ plunging from rest into a
Schwarzschild BH (see \cite{Oohara:1984bw} for details and notation). The
dominant (quadrupolar) mode of the radiation shows a ``bump,'' as expected, at
$\omega=2\Omega_{\rm mb}=(4M)^{-1}$. Since $E_{\rm tot}\sim -\log \delta$ as
$\delta\to 0$,
%and the dominant (quadrupolar) mode of the radiation shows a ``bump,'' as
%expected, at $\omega=2\Omega_{\rm mb}=(4M)^{-1}$. 
radiative effects cannot be neglected in the analysis.
%, and once again they contribute to reduce the BSW effect.

In conclusion, frame dragging effects in Kerr BHs can {\it in principle}
accelerate particle to high energies, but (1) astrophysical restrictions on the spin
severely limit the maximum CM energies in the collisions; (2) just as
radiative losses constrain the performance of particle accelerators,
gravitational radiation and back-reaction constrain the maximum CM energy for collisions around
Kerr BHs; (3) the exponential sensitivity of whirling orbits to initial
conditions requires significant fine-tuning to get sensible cross sections for
the highest-energy collisions.
%, and even then back-reaction effects seem to
%make such collisions extremely unlikely.
%Moreoever, unlike in accelerators, one cannot just pump in more energy or
%adjust the magnitude and frequency of the accelerating electric
%fields. Radiation loss effects have to be taken into account in the
%analysis. We have not ruled out the existence of geodesics giving rise to
%large CM energies. In fact there are an infinite number of such geodesics,
%some of which are closed, zoom-whirl orbits. However, the point of this
%comment is to show that any such discussion must carefully take into account
%gravitational-wave emission in the process.

%In the words of Eric Poisson, ``one can do better by abandoning the dangerous
%fiction of a point particle'' \cite{Poisson:2009dm}.

%%%%%%%%%%%%%%%%%%%%%%%%%%%%%%%%%%%%%%%%%%%%%%%%%%%%%%%%%%%%%%%%%%%%%%%%%%%%%%
{\bf \em Acknowledgements.}
%%%%%%%%%%%%%%%%%%%%%%%%%%%%%%%%%%%%%%%%%%%%%%%%%%%%%%%%%%%%%%%%%%%%%%%%%%%%%%
We thank M.~Ba\~nados, J.~Silk and S.~West for useful comments. 
This work was partially supported by NSF grants PHY-0900735, PHY-0745779,
PHY-0601459, PHY-0652995, PHY-090003, FCT PTDC/FIS/64175/2006,
PTDC/FIS/098025/2008, PTDC/FIS/098032/2008, the Alfred P. Sloan Foundation and
the Sherman Fairchild Foundation.

%%%%%%%%%%%%%%%%%%%%%%%%%%%%%%%%%%%%%%%%%%%%%%%%%%%%%%%%%%%%%%%%%%%%%%%%%%%%%%


\begin{thebibliography}{99}

%\cite{Banados:2009pr}
\bibitem{Banados:2009pr}
  M.~Banados, J.~Silk and S.~M.~West,
  %``Kerr Black Holes as Particle Accelerators to Arbitrarily High Energy,''
  Phys. Rev. Lett. {\bf 103}, 111102 (2009);
  arXiv:0909.0169 [hep-ph].
  %%CITATION = ARXIV:0909.0169;%%

%\cite{Thorne:1974ve}
\bibitem{Thorne:1974ve}
  K.~S.~Thorne,
  %``Disk Accretion Onto A Black Hole. 2. Evolution Of The Hole,''
  Astrophys.\ J.\  {\bf 191}, 507 (1974).
  %%CITATION = ASJOA,191,507;%%

%\cite{Oohara:1984bw}
\bibitem{Oohara:1984bw}
  K.~I.~Oohara and T.~Nakamura,
  %``Energy, Momentum And Angular Momentum Of Gravitational Waves Induced By A
  %Particle Plunging Into A Schwarzschild Black Hole,''
  Prog.\ Theor.\ Phys.\  {\bf 70}, 757 (1983).
  
\end{thebibliography}
\end{document}